\documentclass[prb,twocolumn,showpacs,preprintnumbers,amsmath,amssymb]{revtex4}
\usepackage{graphicx}

\begin{document}

\title{Energies of the first row atoms from quantum Monte Carlo}

\author{M.D.\ Brown}
\author{J.R.\ Trail}
\author{P.\ L\'{o}pez R\'{i}os}
\author{R.J.\ Needs}
\affiliation{Theory of Condensed Matter Group, Cavendish Laboratory,
J.J. Thomson Avenue, Cambridge, CB3 0HE, UK}

\date{June, 2007}

\begin{abstract}
All-electron variational and diffusion quantum Monte Carlo
calculations of the ground state energies of the first row atoms (Li
to Ne) are reported. We use trial wavefunctions of four types: single
determinant Slater-Jastrow wavefunctions; multi-determinant
Slater-Jastrow wavefunctions; single determinant Slater-Jastrow
wavefunctions with backflow transformations; multi-determinant
Slater-Jastrow wavefunctions with backflow transformations.  At the
diffusion quantum Monte Carlo level and using our best trial
wavefunctions we recover 99\% or more of the correlation energy for
Li, Be, B, C, N, and Ne, 97\% for O, and 98\% for F.  
\end{abstract}


\maketitle

\section{Introduction}
\label{intro}

Quantum Monte Carlo (QMC) methods can yield highly accurate energies
for atoms, molecules and solids \cite{RMP_2001,NATO_book},
offering an alternative to quantum chemistry approaches such as
Configuration Interaction and Coupled Cluster \cite{Szabo_Ostlund}.
Although the cost of QMC calculations is large, the scaling with
system size is much better than for accurate quantum chemistry
approaches, and QMC methods have been applied to systems with of order
10$^3$ electrons \cite{Maezono}.

The two most commonly used continuum QMC methods are variational
quantum Monte Carlo (VMC) and diffusion quantum Monte Carlo
(DMC). Stochastic integration is used in VMC to evaluate expectation
values with a trial wavefunction, to a precision which can be
systematically improved by increasing the number of configurations
sampled.  DMC is a stochastic projector method in which a wavefunction
is evolved in imaginary time towards the ground state. Exact
projector methods suffer from a fermion sign problem, and to overcome
this we use the standard fixed-node approximation
\cite{Anderson,Ceperley_Alder} in which the nodal surface of
the wavefunction is fixed during the evolution.  The fixed-node
approximation is uncontrolled, but the DMC energy is lower than the
VMC energy with the same input trial wavefunction, and both give upper
bounds on the exact energy.

DMC gives the lowest energy consistent with the fixed nodal surface.
It is therefore advantageous to optimize parameters in the trial
wavefunction, which may improve the trial nodal surface.  The simplest
and most widely-used trial wavefunction is the Slater-Jastrow (SJ)
form, which consists of a Slater determinant of single-particle
orbitals multiplied by a Jastrow correlation factor. More advanced
wavefunctions can be obtained by, for example, replacing the single
determinant by a sum over configuration state functions (CSFs), by
using backflow transformations of the electronic coordinates,
\cite{Pablo_BF,Neil_Ne,Feynman,Feynman_Cohen}, or
by using pairing wavefunctions
\cite{Casula03,Casula04,Bajdich}.

We present results of QMC calculations for the first row atoms (Li to
Ne) obtained using trial wavefunctions of single and multi-determinant
forms constructed from orbitals calculated on radial grids, both with
and without backflow transformations.  Our aims are to assess the
different roles of multi-determinants and backflow in improving the
trial wavefunctions for the first row atoms and to obtain benchmark
VMC and DMC energies for them. We find that the use of multiple CSFs
and backflow transformations recovers a substantial additional fraction
of the correlation energy at both the VMC and DMC levels.  Our best
wavefunctions recover 99\% or more of the correlation energy at the
DMC level for Li, Be, B, C, N, and Ne, and more than 97\% for O and
98\% for F. This represents a substantial advance upon previous
DMC calculations.

\section{Wavefunction form and optimization}
\label{form_and_opt}

Our multi-CSF SJ wavefunctions with backflow transformations can be
written in the form
\begin{equation}
\Psi(\{{\bf r}_{i}\},{\bf p}) =
e^{J(\{{\bf r}_{i}\},{\bf b})}\sum_{n=1}^{N_{\rm
CSF}}\alpha_{n}
\Phi_{n}( \{{\bf x}_{i}\} ),
\label{eq:1}
\end{equation}
where each ${\bf x}_{i} = {\bf x}_{i} (\{{\bf r}_{i}\}, {\bf c})$
results from the backflow transformation of the electronic
coordinates $\{{\bf r}_{i}\}$,
each $\Phi_{n}$ is a multideterminant representation of a CSF, and
${\bf p}=({\bf a},{\bf b},{\bf c})$ are the variable
wavefunction parameters: the CSF coefficients ${\bf a}$, the
Jastrow parameters ${\bf b}$, and the backflow parameters
${\bf c}$.  It is worth noting that the Jastrow factor normally
introduces spin-contamination \cite{huang98} and the backflow
transformation angular momentum contamination into the trial
wavefunction, so that $\Psi$ will not normally be a spin/spatial
angular momentum eigenstate.  However, the VMC and DMC energies still
provide upper bounds on the true ground state energies.

The general form of our Jastrow factor
$J(\{{\bf r}_{i}\},{\bf b})$ is described in an earlier paper
\cite{Neil_J}. In this study we used Jastrow factors containing
electron-electron, electron-nucleus, and electron-electron-nucleus
terms. Our backflow transformations also consisted of
electron-electron, electron-nucleus, and electron-electron-nucleus
terms, as described in Ref.\ \cite{Pablo_BF}. For both the Jastrow and
backflow functions, the expansion orders and spin dependencies were
chosen by increasing the variational freedom until no further lowering
of the VMC energy was observed.  The maximum number of variable
parameters in the trial wavefunctions varied from 146 for Li to 252
for B, N, C, O, and F.  The number of variable parameters is therefore
only weakly dependent on the atomic number.

We used the multiconfigurational Hartree-Fock (MCHF) atomic-structure
package \textsc{ATSP2K} \cite{fischer07} to construct the $\Phi_{n}$
functions appearing in Eq.~(\ref{eq:1}).  This provides the lowest
energy solution of the many-electron problem where the solution is
limited to an `active space' (AS) specified by an allowed set of
subshell occupation numbers, or configurations.
The wavefunction may formally be written as a sum of CSFs,
\begin{equation}
| LS \rangle_{\rm AS} = \sum_{\nu}^{\rm AS} a_{\nu} |\nu LS
  \rangle,
\end{equation}
where $\nu$ specifies a configuration (for example, $1s^22s^22p^3$)
and seniority number \cite{fischer97}.  Each CSF, $|\nu LS \rangle$,
is defined by a radial orbital for each subshell, and by the total
spatial/spin angular momentum quantum numbers, usually denoted
$^{2S+1}L$.

An additional freedom exists because the underlying symmetry of the
Hamiltonian results in degenerate solutions corresponding to different
projection angular momentum eigenvalues, $L_z$ and $S_z$.  These
additional quantum numbers are not required to specify the total
energy, the coefficients $a_{\nu}$, and the mean field equations
for the radial orbitals, and they are not provided by the
\textsc{ATSP2K} code.  As a result it is not straightforward to obtain
an explicit form for the CSFs which is suitable for use in QMC
calculations.  For the calculations presented here an appropriate
representation of each CSF, $\Phi_{n}$, was constructed as follows.

First we note that two separate angular momentum eigenstates
containing $n_1$ and $n_2$ electrons may be coupled together to
generate angular momentum eigenstates containing $n_1+n_2$ electrons
\cite{fischer97}.  Starting with single electrons and performing such
a coupling iteratively for every allowed combination provides all of
the total spin/angular momentum eigenstates possible for a given
configuration as sums of determinants of spin-orbitals.  Due care was
taken to avoid redundant computation, to respect antisymmetry, and to
employ the same coupling order/sign conventions \cite{gaigalas98} as
used in \textsc{ATSP2K}, allowing the efficient construction of each
CSF as a sum of determinants of one-electron spin-orbitals.

A different wavefunction results from each choice of $L_z$ and $S_z$.
The spin independence of the Hamiltonian allows a direct
conversion of the sum of determinants of complex spin-orbitals into a
sum of products of two determinants of real orbitals, one for spin up
and one for spin down electrons.  This procedure provides a
representation of each CSF in the form
\begin{equation}
\Phi_{\nu L S} = \sum_j a_j D_j^{\uparrow}( \nu L S )
D_j^{\downarrow}( \nu L S ),
\end{equation}
where the determinants are of real single-particle orbitals, and such
that $\sum_{\nu} a_v \Phi_{\nu L S}$ has the same energy
expectation value as the original MCHF wavefunction.  Note that
although $\Phi_{\nu L S}$ has total spatial angular momentum $L$ and
projected spin angular momentum $S_z$, it is not necessarily an
eigenstate of $L_z$ or $S^2$.

\begin{figure}[tbp]
\begin{center}
\includegraphics[width=0.48\textwidth]{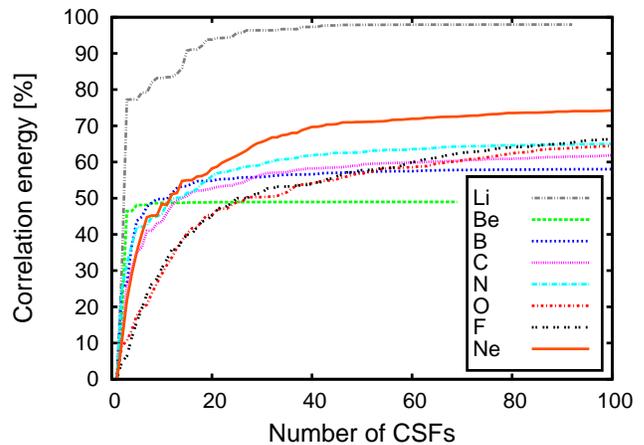}
\end{center}
\caption{(Color online) The percentage of the correlation energy
recovered for each atom as a function of the number of CSFs included
in the MCHF wavefunction. Core excitations were included for Li but
not for the other atoms.}
\label{fig:csfs}
\end{figure}

It should be clear that the $\Phi_{n}$ are not uniquely specified by
($\nu,^{2S+1}L$) due to the spherical symmetry of the underlying
Hamiltonian.  In what follows we make the specific choice $|L_z|=L$,
as this provides the smallest multideterminant expansions, and
$|S_z|=S$, as this provides that same number of spin up/down electrons
as predicted by Hund's first rule.

\begin{table*}[tp]
\begin{tabular}{l|r@{.}lr@{.}lr@{.}lr@{.}lr@{.}lr@{.}lr@{.}lr@{.}l}
\hline
\hline
& \multicolumn{2}{c}{Li ($^{2}$S)} & \multicolumn{2}{c}{Be ($^{1}$S)} &
\multicolumn{2}{c}{B ($^{2}$P)} &
\multicolumn{2}{c}{C ($^{3}$P)} & \multicolumn{2}{c}{N ($^{4}$S)} &
\multicolumn{2}{c}{O ($^{3}$P)} & \multicolumn{2}{c}{F ($^{2}$P)} &
\multicolumn{2}{c}{Ne ($^{1}$S)} \\ \hline
VMC &
$  -7$&$47683(3)$ &
$  -14$&$6311(1)$ &
$  -24$&$6056(2)$ &
$  -37$&$8147(1)$ &
$  -54$&$5482(3)$ &
$  -75$&$0233(3)$ &
$  -99$&$6874(3)$ &
$  -128$&$9047(4)$ \\
VMC-BF &
$  -7$&$47792(2)$ &
$  -14$&$6426(1)$ &
$  -24$&$6377(1)$ &
$  -37$&$8315(1)$ &
$  -54$&$5776(1)$ &
$  -75$&$0522(2)$ &
$  -99$&$7187(2)$ &
$  -128$&$9263(3)$ \\
VMC-MD &
$  -7$&$47752(3)$ &
$  -14$&$66630(4)$ &
$  -24$&$65055(6)$ &
$  -37$&$8383(1)$ &
$  -54$&$5782(1)$ &
$  -75$&$0429(2)$ &
$  -99$&$7054(3$ &
$  -128$&$9206(3)$ \\
VMC-MDBF &
  $  -7$&$47799(1)$ &
  $ -14$&$66716(2)$ &
  $ -24$&$65254(4)$ &
  $ -37$&$84130(8)$ &
  $ -54$&$5840(1) $ &
  $ -75$&$0566(2) $ &
  $ -99$&$7220(2) $ &
  $ -128$&$9246(4)$ \\ \hline
DMC &
$  -7$&$47802(1)$ &
$  -14$&$65717(7)$ &
$  -24$&$63978(5)$ &
$  -37$&$8295(1)$ &
$  -54$&$5769(4)$ &
$  -75$&$0516(5)$ &
$  -99$&$7167(8)$ &
$  -128$&$9223(4)$ \\
DMC-BF &
$  -7$&$478042(4)$ &
$  -14$&$65811(6)$ &
$  -24$&$64548(9)$ &
$  -37$&$83709(8)$ &
$  -54$&$5827(2)$ &
$  -75$&$0584(2)$ &
$  -99$&$7248(2)$ &
$  -128$&$9321(2)$ \\
DMC-MD &
$  -7$&$47800(1)$ &
$  -14$&$66729(1)$ &
$  -24$&$65325(5)$ &
$  -37$&$84317(7)$ &
$  -54$&$5857(2)$ &
$  -75$&$0578(3)$ &
$  -99$&$7237(3)$ &
$  -128$&$9307(3)$ \\
DMC-MDBF &
  $   -7$&$478060(2)$ &
  $  -14$&$667328(6)$ &
  $  -24$&$65362(3)$  &
  $  -37$&$84388(5)$  &
  $  -54$&$5873(1)$   &
  $  -75$&$0615(2)$   &
  $  -99$&$7277(2)$   &
  $ -128$&$9339(3)$ \\
DMC2-MDBF &
  $   -7$&$478058(2)$ &
  $  -14$&$667319(6)$ &
  $  -24$&$65357(3)$  &
  $  -37$&$84385(4)$  &
  $  -54$&$5873(1)$   &
  $  -75$&$0617(2)$   &
  $  -99$&$7274(2)$   &
  $  -128$&$9339(3)$ \\ \hline
  $E_{\rm HF}$ &
  $  -7$&$432727$ &
  $ -14$&$573023$ &
  $ -24$&$529061$ &
  $ -37$&$688619$ &
  $ -54$&$400934$ &
  $ -74$&$809398$ &
  $ -99$&$409349$ &
  $-128$&$547098$ \\
  $E_{\rm exact}$&
  $  -7$&$47806032^{a}$ &
  $ -14$&$66736^{b}$ &
  $ -24$&$65391^{b}$ &
  $ -37$&$8450^{b}$ &
  $ -54$&$5892^{b}$ &
  $ -75$&$0673^{b}$ &
  $ -99$&$7339^{b}$ &
  $-128$&$9376^{b}$ \\
  $E_{\rm HF}-E_{\rm exact}$ &
  $ 0$&$04533332 $ &
  $ 0$&$094337   $ &
  $ 0$&$124849   $ &
  $ 0$&$156381   $ &
  $ 0$&$188266   $ &
  $ 0$&$257902   $ &
  $ 0$&$324551   $ &
  $ 0$&$390502   $ \\
DMC2-corr\% &
  $ 99$&$995(4)\%$ &
  $ 99$&$957(6)\%$ &
  $ 99$&$73(2)\%$ &
  $ 99$&$26(3)\%$ &
  $ 98$&$99(5)\%$ &
  $ 97$&$83(8)\%$ &
  $ 98$&$00(6)\%$ &
  $ 99$&$05(8)\%$ \\
\hline
\hline
\end{tabular}
\caption{The VMC and DMC energies (in Hartrees) for each atom, using a
single determinant SJ wavefunction, a single determinant SJ
wavefunction with backflow (BF), multiple determinants (MD), and both
together (MDBF). The DMC2-MDBF results are for a second DMC run with
half the timestep of DMC-MDBF.  Also shown are the Hartree-Fock single
determinant energies ($E_{\rm HF}$), the ``exact'' energies $E_{\rm exact}$,
the correlation energies $E_{\rm HF}-E_{\rm exact}$, and the percentage of
the correlation energy recovered by the DMC2-MDBF calculations
(DMC2-corr\%). The numbers in parentheses indicate the statistical
uncertainty in the last digit shown. \newline
$^{a}$ taken from Ref.~\cite{Puchalski06} (rounded to 9 significant
figures). \newline
$^{b}$ taken from Ref.~\cite{chakravorty93}.}
\label{tab:all}
\end{table*}

To define the multideterminant part of Eq.~(\ref{eq:1}) we take
$^{2S+1}L$ from the lowest energy HF state, and define an AS using the
following rules for allowed excitations from the HF ground state
configuration: only single and double (SD) electronic excitations were 
allowed, excitations of the $1s$ core electrons were not allowed, and 
excitations were allowed into orbitals with quantum numbers $n \leq 7$ 
and $l \leq 5$.  A MCHF calculation was
then performed for the chosen AS, and the resulting CSFs ordered by
weight, $a_v^2$.  All but the $N_{CSF}$ largest-weight configurations
were then discarded, and a MCHF calculation was performed using this
reduced AS$(N_{CSF})$ active space.  The percentages of the correlation
energies recovered at the MCHF level are shown in Fig.~\ref{fig:csfs}
for $N_{CSF} \leq 100$.

Li is an exception: the above rules define an active space that is 
equivalent to HF because of the absence of $1s$ excitations. In order 
to generate a nontrivial AS for Li we allowed excitations of all 
electrons into orbitals with quantum numbers $n\leq5$ and $l\leq5$.

We chose $N_{CSF}=20$ for the multideterminant expansions used
in the QMC calculations (corresponding to a number of determinants 
ranging from 83 for Li to 499 for Ne).  Although the application of this
selection criterion to the first row atoms is necessarily somewhat
arbitrary, it is not without justification. Figure~\ref{fig:csfs}
shows that for around 20 CSFs the fraction of correlation energy
recovered within MCHF is roughly equivalent from Be to Ne. Including 
further orbitals in the initial `large' AS, including triple 
excitations, or allowing core excitations, did not significantly change 
the MCHF energies for $N_{CSF} \leq 20$, with the exceptions of Li and 
Be for which core excitations improved the MCHF energies significantly 
for small $N_{CSF}$, but had a negligible impact on the DMC results.

\begin{figure}[b]
\includegraphics[width=0.5\textwidth]{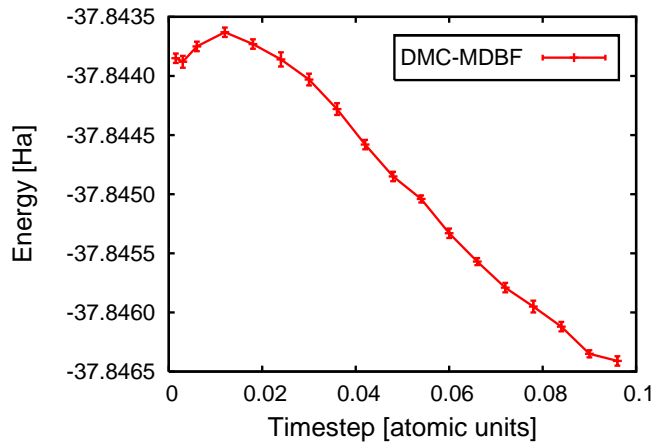}
\caption{(Color online) The variation of the DMC energy (in Hartrees)
with timestep for the C atom, using a MDBF wavefunction. The
statistical error bars on the data are shown.}
\label{fig:cdt}
\end{figure}

The variable parameters $\bf p$ were optimized by minimizing the VMC
energy, using
the scheme developed very recently by Umrigar \emph{et al.}
\cite{Umrigar_emin,Toulouse_emin}.  This scheme is a
generalisation of the method of Nightingale and Melik-Alaverdian
\cite{N_MA} to include non-linear parameters in the trial
wavefunction. In the past, variance minimization techniques were 
normally used because they are robust for optimizing Jastrow 
factors, but in fact they have been found to behave poorly when 
optimizing linear coefficients.  The method of Nightingale and 
Melik-Alaverdian \cite{N_MA} is very robust for optimizing linear 
coefficients, and Umrigar \emph{et al.} 
\cite{Umrigar_emin,Toulouse_emin} have developed an energy
minimization scheme which is robust for both linear and non-linear
parameters.  All of the wavefunction parameters were optimized
simultaneously in our calculations.  In some cases we obtained
significantly lower VMC and DMC energies with energy minimization 
than with variance minimization when optimizing the CSF coefficients
${\bf a}$, and the performance for the Jastrow and backflow 
parameters (${\bf b}$ and ${\bf c}$) was also slightly improved.

\section{Results}
\label{results}

We used the \textsc{casino} code \cite{CASINO} for all of our VMC and
DMC calculations.  The wavefunction optimizations were performed using
$10^{5}$ statistically independent electronic configurations.  The
target population of walkers in the DMC calculations was about 2000 in
each case, which should ensure that any population control bias is
negligible.  The timesteps for the DMC calculations were chosen to
make the systematic finite timestep errors smaller than the random
statistical errors.  The smallest timesteps used, corresponding to the
DMC2 data in Table~\ref{tab:all}, ranged from 0.00375~a.u.\ for Li to
0.0007~a.u.\ for Ne.  The variation of the DMC energy of the C atom
with timestep is shown in Fig.~\ref{fig:cdt}, and the shape of this
curve is typical of our calculations for the first row atoms.
For C, the DMC acceptance ratios varied from 93(1)\% for the largets 
time step (0.096 a.u.) to 99.8(1)\% for the smallest time step 
(0.0015 a.u.).

\begin{figure}[t]
\includegraphics[width=0.5\textwidth]{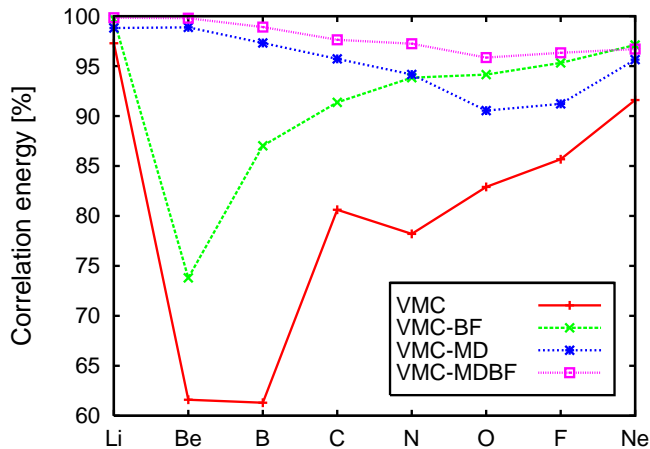}
\caption{(Color online) The percentage of the correlation energy
recovered for each atom within VMC, using a single determinant
SJ wavefunction (VMC) with the addition of backflow
(VMC-BF), multiple determinants (VMC-MD), and both together
(VMC-MDBF). The statistical error bars (not shown) are smaller than
the symbols.}
\label{fig:vmc}
\end{figure}

\begin{figure}[b]
\includegraphics[width=0.5\textwidth]{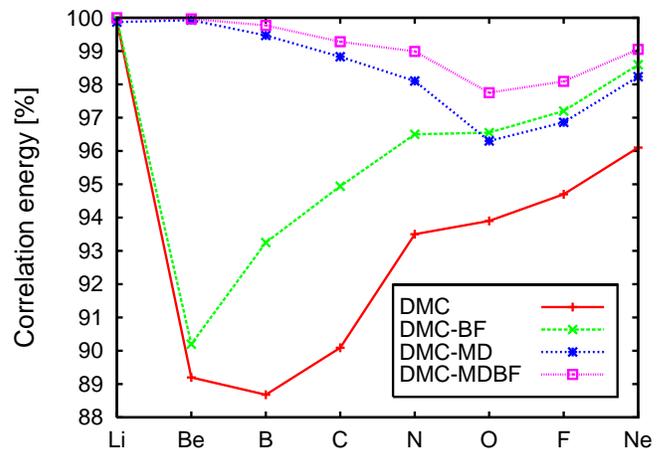}
\caption{(Color online) The percentage of the correlation energy
recovered for each atom within DMC, using a single determinant
SJ wavefunction (DMC) with the addition of backflow
(DMC-BF), multiple determinants (DMC-MD), and both together
(DMC-MDBF). The statistical error bars (not shown) are smaller than
the symbols.}
\label{fig:dmc}
\end{figure}

Table~\ref{tab:all} gives the VMC and DMC energies, including the most
accurate
correlation energy percentages achieved for each atom.  The ``exact''
non-relativistic energies, assuming a point nucleus of infinite mass,
were taken from Refs.\ \cite{Puchalski06,chakravorty93}.
Fig.~\ref{fig:vmc} shows the percentage of each atom's correlation
energy recovered by the optimized wavefunctions within VMC as the
multi-CSF expansion and backflow transformations were added.
Fig.~\ref{fig:dmc} gives the same information at the DMC level.

The general shapes of the curves in Figs.~\ref{fig:vmc} and
\ref{fig:dmc} are similar, but note that the VMC graph covers a range
of 40\%, while the DMC one covers only 12\%.  Our best VMC results
surpass the accuracy previously achieved in DMC calculations using
single determinant SJ trial wavefunctions 
\cite{luchow96,Langfelder_JCP,Hongo_MatTrans}. Our single determinant
DMC results are similar to those reported by other authors
\cite{luchow96,Langfelder_JCP,Hongo_MatTrans}.  For Li
and Be our best DMC energies are within error bars of some previous DMC
calculations, but for the other atoms our results are clearly
superior.  Flad \emph{et al.} \cite{flad97} performed multideterminant
DMC calculations for the atoms B to F, but we have used larger
expansions and our results are already superior at the DMC-MD level.
Casula and Sorella \cite{Casula03} performed DMC calculations for the
atoms Li-Ne with correlated geminal pairing wavefunctions, but our
energies are substantially better than theirs.  VMC and DMC
calculations using backflow wavefunctions have been reported for Li
\cite{Pablo_BF} and Ne \cite{Neil_Ne}.  L\"uchow and Fink
\cite{luchow00} performed a DMC calculation for N using a pair natural
orbital configuration interaction trial wavefunction, retrieving
98.2\% of the correlation energy, which is very similar to our DMC-MD
result, but our DMC-MDBF energy is lower.  Our results for Li and Be
are surpassed by Hylleraas variational calculations
\cite{Puchalski06,komasa02}.  We should also note the existence
of high quality non-variational results such as those from explicitly
correlated $r_{12}$ Configuration Interaction and Coupled Cluster
methods \cite{gdanitz98,noga01}.

Each of the VMC and DMC energies for Li corresponds to recovering well
over 90\% of the correlation energy.  The single-determinant HF nodal
surface of Li is extremely accurate, but incorporating backflow
improves the DMC energy noticeably.  Backflow gives a significant
improvement for Be at the VMC level, but a very modest improvement
within DMC.  The ground state wavefunction for Be contains a
substantial admixture of the $1s^2 2p^2$ excited state configuration
which significantly alters the nodal surface.  The single determinant
$1s^2 2s^2$ HF nodal surface divides the configuration space into four
nodal pockets, while the exact wavefunction has only two nodal pockets
\cite{Glauser,Bressanini}.  It does not appear possible to
achieve an accurate DMC energy for Be with a trial wavefunction
containing four nodal pockets.  The continuous backflow
transformations we use are incapable of changing the number of nodal
pockets \cite{Pablo_BF} and consequently the single determinant DMC
energy for Be is very poor.  When an appropriate amount of the $1s^2
2p^2$ configuration is included in the trial wavefunction, the DMC
energy improves dramatically, and backflow contributes a small
additional improvement.

The percentages of the correlation energy recovered for N at the VMC
level from the single determinant backflow and 20 CSF wavefunctions
are almost the same, while for higher atomic numbers backflow gives
more correlation energy than the CSFs.  Fig.~\ref{fig:csfs} shows that
the percentages of the correlation energy recovered within the MCHF
calculations for O and N increase relatively slowly with the number of
CSFs, which may be connected with the lower percentages of correlation
energy recovered in the MD and MDBF calculations for these atoms.
Note that, apart from Li and Be where the DMC-MD energies with 20 CSFs
are already excellent, backflow adds substantially to the correlation
energy recovered.  It seems that backflow and multi-CSFs improve the
single determinant wavefunctions in rather different ways, and
therefore it can be very advantageous to combine them.

One might expect that the VMC-MDBF energies would always be lower than
the VMC-BF and VMC-MD energies, as the wavefunction parameters are
obtained by minimizing the VMC energy.  Table~\ref{tab:all} shows,
however, that this does not hold for Ne.  The most likely reason for
this is that the HF and MCHF orbitals are different, so that the
single determinant wavefunction does not correspond to a particular
choice of the MD wavefunction parameter values.  This also indicates
that further reductions in the VMC energy could be obtained by
optimizing the orbitals along with the Jastrow and backflow parameters
and the CSF coefficients.

\section{Discussion}
\label{disc}

As can be seen from Fig.~\ref{fig:dmc}, we recover 99\% or more of the
correlation energy for all the atoms except O and F.  The different
freedoms introduced by the multi-CSF expansion and backflow
transformations are both vital in achieving the best energies,
indicating that they capture different aspects of the wavefunction
correlation.  Backflow appears to describe parts of the dynamic
correlation beyond the SJ wavefunction while the low-energy CSFs
beyond the HF ground state configuration describe static correlation
energy.  Our results contribute to the growing evidence
\cite{Pablo_BF,Neil_Ne} that adding backflow transformations to
SJ wavefunctions can significantly improve the nodal
surface for systems from atoms to solids.

It should be noted that, while it might be difficult to substantially
improve the Jastrow factors and backflow transformations, we could
easily include many more CSFs in the trial wavefunctions without a
prohibitive increase in computational expense.  The orbitals in the
CSFs could also be optimized within VMC. The energies reported here do
not therefore represent the current practical limits of the VMC and
DMC methods for all-electron atoms.

For general systems, the number of CSFs required to obtain a 
significant improvement in the energy increases rapidly with the 
number of electrons $N$. The number of required parameters in the 
backflow functions only increases (roughly linearly) with the number 
of inequivalent aoms.  The cost
of moving an electron in a VMC/DMC calculation scales approximately as
$N^2$.  In a VMC-BF/DMC-BF calculation the cost scales as $N^3$
because all of the orbitals must be evaluated at the positions
$\{{\bf x}_{i}\}$ when an electron is moved.  The cost of a VMC-MD/DMC-MD
calculation scales as $N^2M$, where $M$ is the number of determinants,
and the cost of a VMC-MDBF/DMC-MDBF calculation scales as $N^3M$.  The
number of moves required to obtain a fixed statistical error bar in
the energy is smaller for a more accurate trial wavefunction, so 
that the pre-factors in the cost for the more sophisticated wavefunction
forms are smaller. As an example, for the C atom the relative 
computational costs to achieve the same error bar are 1.0/3.6/6.2/17.8 
for the VMC SD/SDBF/MD/MDBF calculations, while in DMC the corresponding 
cost ratios are 9.0/13.7/15.1/31.9. Overall, we expect that we can 
maintain the level of accuracy achieved here for small molecules.

We obtained superior results with the energy minimization scheme of
Umrigar \emph{et al.} \cite{Umrigar_emin} than with variance
minimization procedures
\cite{Umrigar_varmin,Kent_varmin,Neil_varminlinjas}.
The energy minimization method is particularly advantageous when
multiple determinants are used because it is well-suited to optimizing
linear parameters, whereas variance minimization encounters some
difficulties in optimizing CSF coefficients.  The combination of
energy minimization and the MD and BF wavefunction forms has led to
substantial reductions in the VMC energies, which are very important
because the variance of the DMC energy has been found to be
proportional to the error in the VMC energy \cite{Ceperley,Ma}.
Although reductions in the DMC energies from the improved trial
wavefunctions are smaller than in VMC, they represent substantial
fractions of the remaining correlation energy. Our best DMC
energies for Li and Be are within error bars of previously published
DMC energies, but our results for the higher atomic number atoms are
markedly superior.

\begin{acknowledgments}
Financial support was provided by the Engineering and Physical
Sciences Research Council (EPSRC) of the UK. Computing resources were
provided by the University of Cambridge High Performance Computing
Service (HPCS).
\end{acknowledgments}


\begin{thebibliography}{00}

\bibitem{RMP_2001}
W.M.C.\ Foulkes, L.\ Mitas, R.J.\ Needs, and G.\ Rajagopal, Rev.\ Mod.\
Phys.\ {\bf 73}, 33 (2001).

\bibitem{NATO_book}
C.J.\ Umrigar and M.P.\ Nightingale, eds., \emph{Quantum Monte Carlo Methods in
Physics and Chemistry, NATO ASI Ser.\ C 525} (Kluwer, Dordrecht, 1999).

\bibitem{Szabo_Ostlund}
A.\ Szabo and N.S.\ Ostlund, \emph{Modern Quantum Chemistry} (Dover, 1996).

\bibitem{Maezono} 
R.\ Maezono, A.\ Ma, M.D.\ Towler and R.J.\ Needs, Phys.\ Rev.\
Lett.\ {\bf 98}, 025701 (2007).

\bibitem{Anderson} 
J.B.\ Anderson, J.\ Chem.\ Phys.\ {\bf 63}, 1499 (1975).

\bibitem{Ceperley_Alder}
D.M.\ Ceperley and B.J.\ Alder, Phys.\ Rev.\ Lett.\ {\bf 45}, 566 (1980).

\bibitem{Pablo_BF}
P.\ L\'{o}pez R\'{i}os, A.\ Ma, N.D.\ Drummond, M.D.\ Towler and R.J.\
Needs, Phys.\ Rev.\ E {\bf 74}, 066701 (2006).

\bibitem{Neil_Ne}
N.D.\ Drummond, P.\ L\'{o}pez R\'{i}os, A.\ Ma, J.R.\ Trail, G.G.\
Spink, M.D.\ Towler and R.J.\ Needs, J.\ Chem.\ Phys.\ {\bf 124}, 224104 (2006).

\bibitem{Feynman}
R.P.\ Feynman, Phys.\ Rev.\ {\bf 94}, 262 (1954).

\bibitem{Feynman_Cohen}
R.P.\ Feynman and M.\ Cohen, Phys.\ Rev.\ {\bf 102}, 1189 (1956).

\bibitem{Casula03} 
M.\ Casula and S.\ Sorella, J.\ Chem.\ Phys.\ {\bf 119}, 6500 (2003).

\bibitem{Casula04} 
M.\ Casula, C.\ Attaccalite and S.\ Sorella, J.\ Chem.\ Phys.\ {\bf 121},
7110 (2004).

\bibitem{Bajdich}
M.\ Bajdich, L.\ Mitas, G.\ Drobn\'y, L.K.\ Wagner and K.E.\
Schmidt, Phys. Rev. Lett. {\bf 96}, 130201 (2006).

\bibitem{huang98}
C.\ Huang, C.\ Filippi and C.J.\ Umrigar, J.\ Chem.\ Phys.\ {\bf 108}, 8838 (1998).

\bibitem{Neil_J}
N.D.\ Drummond, M.D.\ Towler and R.J.\ Needs, Phys.\ Rev.\ B {\bf 70}, 235119 (2004).

\bibitem{fischer07}
C.\ Froese Fischer, G.\ Tachiev, G.\ Gaigalas and M.\ Godefroid 
Comput.\ Phys.\ Commun. {\bf 176}, 559 (2007). (http://atoms.vuse.vanderbilt.edu).

\bibitem{fischer97} 
C.\ Froese Fischer, T.\ Brage and P.\ J\"onsson, \emph{Computational
Atomic Structure: An MCHF approach} (IOP Publishing Ltd.\, 1997).

\bibitem{gaigalas98}
G.\ Gaigalas, Z.\ Rudzikas and C.\ Froese Fischer, At.\ Data\ Nucl.\ Tables\ {\bf 70},
1 (1998).

\bibitem{Umrigar_emin}
C.J.\ Umrigar, J.\ Toulouse, C.\ Filippi, S.\ Sorella and R.G.\ Hennig,
Phys.\ Rev.\ Lett.\ {\bf 98}, 110201 (2007).

\bibitem{Toulouse_emin}
J.\ Toulouse and C.J.\ Umrigar, J.\ Chem.\ Phys.\ {\bf 126}, 084102 (2007).

\bibitem{N_MA}
M.P.\ Nightingale and V.\ Melik-Alaverdian, Phys.\ Rev.\ Lett.\ {\bf 87}, 043401 (2001).

\bibitem{Puchalski06}
M.\ Puchalski and K.\ Pachucki, Phys.\ Rev.\ A {\bf 73}, 022503 (2006).

\bibitem{chakravorty93} 
S.J.\ Chakravorty, S.R.\ Gwaltney, E.R.\ Davidson, F.A.\ Parpia and C.F.\
Fischer, Phys.\ Rev.\ A {\bf 47}, 3649 (1993).

\bibitem{CASINO}
R.J.\ Needs, M.D.\ Towler, N.D.\ Drummond and P.\ L\'{o}pez R\'{i}os,
\emph{CASINO user's guide, version 2.0.0} (2006).

\bibitem{luchow96}
A.\ L\"uchow and J.B.\ Anderson, J.\ Chem.\ Phys.\ {\bf 105}, 7573 (1996).

\bibitem{Langfelder_JCP}
P.\ Langfelder, S.M.\ Rothstein and J.\ Vrbik, J.\ Chem.\ Phys.\ {\bf 107}, 8525 (1997).

\bibitem{Hongo_MatTrans}
K.\ Hongo, Y.\ Kawazoe and H.\ Yasuhara, Materials Transactions {\bf 47}, 2612 (2006).

\bibitem{flad97}
H.\ Flad, M.\ Caffarel and A.\ Savin, in \emph{Recent Advances in Quantum
Monte Carlo Methods}, edited by W.A.\ Lester, Jr.\ (World Scientific,
Singapore, 1997).

\bibitem{luchow00}
A.\ L\"uchow and R.F.\ Fink, J.\ Chem.\ Phys.\ {\bf 113}, 8457 (2000).

\bibitem{komasa02} 
J.\ Komasa, J.\ Rychlewski and K.\ Jankowski, Phys.\ Rev.\ A {\bf 65}, 042507 (2002).

\bibitem{gdanitz98}
R.J.\ Gdanitz, J.\ Chem.\ Phys.\ {\bf 109}, 9795 (1998).

\bibitem{noga01}
J.\ Noga, P.\ Valiron and W.\ Klopper, J.\ Chem.\ Phys.\ {\bf 115}, 2022 (2001).

\bibitem{Glauser} 
W.A.\ Glauser, W.R.\ Brown, W.A.\ Lester, Jr., D.\ Bressanini, B.L.\
Hammond and M.L.\ Koszykowski, J.\ Chem.\ Phys.\ {\bf 97}, 9200 (1992).

\bibitem{Bressanini} 
D.\ Bressanini, D.M.\ Ceperley and P.J.\ Reynolds, in \emph{Recent
Advances in Quantum Monte Carlo Methods, 2nd ed}.\, edited by W.A.\ Lester, Jr.,
S.M.\ Rothstein and S.\ Tanaka (World Scientific, Singapore, 2002).

\bibitem{Umrigar_varmin}
C.J.\ Umrigar, K.G.\ Wilson and J.W.\ Wilkins, Phys.\ Rev.\ Lett.\ {\bf 60}, 1719 (1988).

\bibitem{Kent_varmin}
P.R.C.\ Kent, R.J.\ Needs and G.\ Rajagopal, Phys.\ Rev.\ B {\bf 59}, 12344 (1999).

\bibitem{Neil_varminlinjas}
N.D.\ Drummond and R.J.\ Needs, Phys.\ Rev.\ B {\bf 72}, 085124 (2005).

\bibitem{Ceperley} 
D.M.\ Ceperley, J.\ Stat.\ Phys.\ {\bf 43}, 815 (1986).

\bibitem{Ma} A.\ Ma, N.D.\ Drummond, M.D.\ Towler and R.J.\ Needs,
Phys.\ Rev.\ E {\bf 71}, 066704 (2005).

\end{thebibliography}
\end{document}